\documentclass[preprint,aps]{revtex4}

\usepackage{graphicx}
\usepackage{graphicx}
\usepackage{epsfig}
\usepackage{amssymb}
\usepackage{graphicx}
\usepackage{dcolumn}
\usepackage{bm}
\newcommand{\ba}{\begin{array}}
\newcommand{\ea}{\end{array}}
\def\br{\begin{eqnarray}}
\def\er{\end{eqnarray}}
\def\be{\begin{equation}}
\def\ee{\end{equation}}

\def\({\left(}
\def\){\right)}

\begin{document}

%
%

\title{Schwinger-Dyson approach and its application to generate a light composite scalar}
\author{A. Doff$^{1}$ and  A. A. Natale$^{2}$}
\affiliation{$^1$Universidade Tecnol\'ogica Federal do Paran\'a - UTFPR - DAFIS
Av Monteiro Lobato Km 04, 84016-210, Ponta Grossa, PR, Brazil \\
agomes@utfpr.edu.br \\
$^2$Instituto de F\'{\i}sica Te\'orica, UNESP  Rua Dr. Bento T. Ferraz, 271, Bloco II, 01140-070, S\~ao Paulo - SP,
Brazil\\
Centro de Ci\^encias Naturais e Humanas, Universidade Federal do ABC, 09210-170, Santo Andr\'e - SP, Brazil\\
natale@ift.unesp.br }







\begin{abstract}
We discuss the possibility of generating a light composite scalar boson, in a scenario that we may generically call Technicolor, or in any variation of a strongly interacting theory, where by light we mean a scalar composite mass about one order of magnitude below the characteristic scale of the strong theory. Instead of most of the studies about a composite Higgs boson, which are based on effective Lagrangians, we  consider this problem in the framework of non-perturbative solutions of the fermionic Schwinger-Dyson and Bethe-Salpeter equations. We study a range of mechanisms proposed during the recent years to form such light composite boson, and verify that such possibility 
seems to be necessarily associated to a fermionic self-energy that decreases slowly with the momentum.

\end{abstract}
\maketitle


\section{Introduction}

\par The $125$ GeV new resonance discovered at the LHC \cite{LHC1,LHC2} has many of the characteristics expected for the Standard Model (SM) Higgs boson. If this particle is a composite or an elementary scalar boson is still an open question that probably will be answered at LHC13 run. The possibility that a light composite state could be discovered at the LHC, in a scenario that we may generically call Technicolor, or in any variation of a strongly interacting theory has been extensively discussed in the literature \cite{sannino}. 

\par The chiral and gauge symmetry breaking in quantum field theories can be promoted by fundamental scalar bosons through the Higgs boson mechanism. However the main ideas about symmetry breaking and spontaneous generation of fermion and gauge boson masses in field theory were
based on the superconductivity theory. Nambu and Jona-Lasinio proposed one of the first field theoretical models based on the ideas of superconductivity, where all the most
important aspects of symmetry breaking and mass generation, as known nowadays, were explored at length \cite{nl}. The model of Ref.\cite{nl}
contains only fermions possessing invariance under chiral symmetry, although this invariance is not respected by the vacuum of the theory
and the fermions acquire a dynamically generated mass ($m_f$). As a consequence of the chiral symmetry breaking by the vacuum the
analysis of the Bethe-Salpeter equation (BSE) shows the presence of Goldstone bosons. These bosons, when the theory is assumed to be
the effective theory of strongly interacting hadrons, are associated to the pions, which are not true Goldstone bosons when the 
constituent fermions have a small bare mass. Besides these aspects Nambu and Jona-Lasinio also verified that the theory presents
a scalar bound state (the sigma meson).   

In Quantum Chromodynamics (QCD) the same mechanism is observed, where the quarks acquire a dynamically generated mass ($m_{dyn}$). This dynamical mass is usually expected to appear as a solution of the Schwinger-Dyson equation (SDE) for the fermion propagator when the coupling constant is above a certain critical value. The condition that implies a dynamical mass for quarks breaking the chiral symmetry is the same one that generates a bound-state massless pion. This happens because the quark propagator SDE is formally the same equation binding a quark and antiquark into the massless pseudoscalar state at zero momentum transfer (the pion).  As shown
by Delbourgo and Scadron \cite{scadron1}, the same similarity of equations happens for the scalar p-wave state of the BSE, indicating the
presence of a bound state with mass $(m_\sigma = 2 m_{dyn})$ and  this scalar meson is the elusive sigma meson \cite{polosa1,polosa2,polosa3}, that is assumed to be the Higgs boson of QCD. 

Inspired in QCD, Technicolor was a theory invented to provide a natural and consistent quantum-field theoretic description of electroweak (EW) symmetry breaking,  without elementary scalar fields. A possibility raised a few years ago is to have a Higgs arising as a composite pseudo-Goldstone boson (PGB) from the strongly interacting sector, where in this case the Higgs mass is protected by an approximate global symmetry and is only generated via quantum effects,  models based on this approach are usually  called composite Higgs models\cite{bella}. As pointed out in Refs.\cite{bella,lane2}, composite Higgs models  require a degree of fine-tuning of parameters and most of the studies about a composite Higgs boson are based on effective Lagrangians \cite{bella}. The approach that we will discuss here is not based on effective Lagrangians or operators, but on the dynamics of the theory, i.e. on the non-perturbative solutions of the Schwinger-Dyson and Bethe-Salpeter equations. The freedom appearing on the coefficients of the many possible operators that can be introduced into an effective Lagrangian in order to describe the SM scalar boson sector, is traded now by the self-energies of the new fermions that form the composite scalar boson. Assuming that the underlying theory is a non-Abelian $SU(N)$ gauge theory, we will verify that the restriction imposed on the theory parameters by the existence of a light scalar composite are quite tight, and the construction of a realistic ``Technicolor" model may indeed be a very precise engineering problem.

In  a very recent paper, ATLAS and CMS Collaborations \cite{ATLAS-CMS}, based on their combined data samples presented an improved precision value for $m_{H} = 125.09 \pm 0.24 GeV$. The improved knowledge of $m_{H}$  yields more precise predictions for the Higgs couplings and until now the coupling strengths to SM particles are consistent within the uncertainties  expected for the SM Higgs boson. Probably in a realistic composite scalar model the fermionic couplings will involve an ETC group(or GUT) with a delicate vacuum alignment between  two type  of composite scalar  bosons $H$ and $H'$,   where one of the composites  resemble a fundamental scalar and  the fermionic masses are not generated as usual, by different ETC mass scales, but  by a different mechanism. In the Ref.\cite{us2} we considered a TC model with  a horizontal symmetry where the top quark (or the third fermionic generation) obtain its mass associated to a large ETC scale, and in this case  only the composite  boson $H'$ is coupled to the third fermionic generation, with a coupling resembling the one of a fundamental
scalar boson, although the aspects of generating the SM fermionic mass spectrum will not be touched here. We just commented this in order to remember that the generation of a light
scalar composite is only part of the problem if we desire the full SM dynamical symmetry breaking, where the generation of the fermionic mass spectra is another enormous step in this
direction, and, probably, the larger part of the problem.

Our intent in this work is not to advocate that the $125$ GeV scalar boson is a composite one, but to discuss how a scalar composite boson can be generated with a mass lighter than the characteristic scale of the strongly interacting theory. Of course, we will resort to our knowledge about QCD, where the lightest scalar boson has a mass about twice the characteristic scale $\Lambda_{QCD}$, as well as we will make analogy
to the SM where the scalar responsible for the gauge symmetry breaking has a mass about one order of magnitude below that of the Fermi scale 
(${\cal{O}}(1)TeV$). 

The advantage of the approach that we shall propose here is that it allows to discover what type of dynamics can lead to a ``light" composite, and also indicates what are the types of effective Lagrangians that favor such composite particle. Indeed we shall discuss that working with
the theory dynamics, i.e. self-energies or bound state solutions, we can restrict the existence of certain terms in the many possible
effective Lagrangians to describe the composite Higgs boson potential.

In this work we  consider the problem of generating  one light composite scalar in a strongly interacting non-Abelian gauge theory assuming a range of mechanisms developed  during the recent years. In Section 2 of this work we study the problem of generating at least one light composite scalar in a theory with a unique characteristic scale $\Lambda$. This analysis will be
performed with the use of the Bethe-Salpeter equation. The same result will also be checked with the use of the effective potential for composite operators in Section 3, and we make a few remarks about
the possible influence of mixing between different scalars formed within the same theory in Section 4. 
In Section 5 we discuss how the mass of a composite Higgs boson is modified in the presence of other interactions, i.e. any interaction that is not the one responsible to form the composite scalar state. In Section 6 we consider the  possibility of generating a light composite scalar happening in the case where we have at least two composite bosons, related to two different
scales and there is a strong mixing between the scalars \cite{ff}, i.e. we may have a see-saw mechanism where one of the scalar composites may turn out to be quite light. In Section 7 we present a brief discussion about how the mixing mechanism, discussed in Section 6, can be  extended to models with more than one TC group. In Section 8 we make a brief remark on the mass of vector composites, because this is one of the main signals of a new strongly interacting theory and the Section 9 contains our conclusions.  

\section{Scalar composite boson mass in isolation: Bethe-Salpeter equation approach}

In this section we shall consider the problem of generating at least one light composite scalar in a strongly interacting non-Abelian gauge theory with a \textit{unique} characteristic scale $\Lambda$. Most of our discussion will be guided by QCD, which is the only strongly interacting non-Abelian theory that we have to compare with. The scale $\Lambda$ has a similar role as the QCD scale ($\Lambda_{QCD}$), where it is known that quarks generate
a condensate
\be
|\left.\left\langle {\bar{q}}q \right\rangle\right. |^{1/3} \approx \Lambda_{QCD} \approx \mu  \, ,
\label{eq1}
\ee
where $\mu$ is the dynamical quark mass.   At the same time that the QCD chiral symmetry is broken Goldstone bosons are formed (the pions) and
a set of scalars are also generated \cite{polosa1,polosa2,polosa3,polosa4}. Considering a minimum number of quarks we shall have at least one light scalar boson (the sigma meson),
whose mass is
\be
m_\sigma = 2 \mu \,\, .
\label{eq2}
\ee

Eq.(\ref{eq2}) comes out from the following relation \cite{scadron1,scadron2}
\be
\Sigma (p^2) \approx  \Phi_{BS}^P (p,q)|_{q \rightarrow 0} \approx \Phi_{BS}^S (p,q)|_{q^2 = 4 \mu^2 }\,\,\, ,
\label{eq3}
\ee
where the solution ($\Sigma (p^2)$) of the fermionic Schwinger-Dyson equation (SDE), that indicates the
generation of a dynamical quark mass and chiral symmetry breaking of QCD, is a solution of 
the homogeneous Bethe-Salpeter equation (BSE) for a massless pseudoscalar bound state ($\Phi_{BS}^P (p,q)|_{q \rightarrow 0}$),
and is also a solution of  the  homogeneous BSE of a scalar p-wave bound state ($\Phi_{BS}^S (p,q)|_{q^2 = 4 \mu^2 }$), which
implies the existence of the scalar (sigma) boson with the mass described above. 

Eq.(\ref{eq2}) has a central point here, and in all subsequent discussion we will verify how this equation can be modified in the general case of a strongly interacting theory. In order to do this we will work with the most general fermionic self-energy or bound state solution, i.e. a solution that may describe any possible dynamics of a $SU(N)$ non-Abelian gauge theory. 

Many models have considered the possibility of a light composite Higgs based on effective Higgs potentials as reviewed in Ref.\cite{bella}. The reason for the existence of the different models (or different potentials) for a composite Higgs, is a consequence of our poor knowledge of the strongly interacting theories; that is reflected in the many possible choices of parameters in the effective potentials. On the other hand the  composite scalar boson mass can be calculated based on the dynamics of the theory \cite{us1}, and this approach, although more complex, is more restrictive than the analysis of potential coefficients in several specific limits.

Our starting point is the most general asymptotic fermionic self-energy expression for a non-Abelian gauge theory \cite{us2,us3}:
\be 
\Sigma (p^2) \sim \mu \left( \frac{ \mu^2 }{p^2}\right)^{\alpha}\left[1 + b g^2 (\mu^2) \ln\left(p^2/\Lambda^2 \right) \right]^{-\gamma\cos (\alpha \pi)}  \,\,\, .
\label{eq4}
\ee	
In the above expression $\Lambda$ is the characteristic mass scale of the strongly interacting theory forming the
composite Higgs boson, and for simplicity we assume $\Lambda \approx \mu$. Note that dynamical mass $\mu$ is not an observable. In principle, it must have a simple relation with $\Lambda$, and from the QCD experience we may expect that they are of the same order. $g$ is the strongly interacting running coupling constant, $b$ is the coefficient of $g^3$ term in the renormalization group $\beta$ function,  
\be
\gamma= 3c/16\pi^2 b,
\label{ga01}
\ee
and  $c$ is the quadratic Casimir operator given by  
\be
c = \frac{1}{2}\left[C_{2}(R_{1}) +  C_{2}(R_{2}) - C_{2}(R_{3})\right],
\label{ca01}
\ee
where $C_{2}(R_{i})$
are the Casimir operators for fermions in the representations  $R_{1}$ and  $R_{2}$ that form a composite boson in the representation $R_{3}$.

The parameter $\alpha$ in Eq.(\ref{eq4}) varies between $0$ and $1$. When $\alpha =1$ Eq.(\ref{eq4}) gives the known asymptotic self-energy behavior predicted by the operator product expansion (OPE) \cite{pol}
\be
\Sigma^{(1)} (p^2 \rightarrow \infty) \sim \frac{ \mu^3 }{p^2} \, .
\label{eq5}
\ee
When $\alpha =0$ we obtain
\be
\Sigma^{(0)} (p^2) \sim \mu \left[1 + b g^2 (\mu^2) \ln\left(p^2/\mu^2 \right) \right]^{-\gamma} \, .
\label{eq6}
\ee  
The asymptotic expression shown in Eq.(\ref{eq6}) was determined in the appendix
of Ref.\cite{cs} and it satisfies the Callan-Symanzik equation.
It has been argued that Eq.(\ref{eq6}) may be a realistic solution in a
scenario where the chiral symmetry breaking is associated to confinement and
the gluons have a dynamically generated mass \cite{us12,us22,us4}. This solution
also appears when using an improved renormalization group approach in QCD,
associated with a finite quark condensate \cite{chan1,chan2,chan3}, and it minimizes the
vacuum energy as long as $n_{f}>5$ \cite{mon}. Moreover, this specific
solution is the only one consistent with Regge-pole like solutions \cite{lan}.
Finally, the apparent explicit breaking of the chiral symmetry described by Eq.(\ref{eq6}) seems
also to be induced by the critical Wilson term in the QCD action \cite{frezz}, which is a quite intriguing result
compatible with arguments presented in Ref.\cite{us12} about the importance of confinement for chiral symmetry breaking.
The important fact is that this is the hardest (in momentum space) asymptotic
behavior allowed for a bound state (or self-energy) solution in a non-Abelian gauge theory, and
it is exactly for this reason that a constraint on $\gamma$ arises from the
BSE normalization condition \cite{lane1,man,les} implying 
\be
\gamma > 1/2 .
\label{eq7}
\ee
This condition has also been re-obtained recently associated with the positivity of the scalar composite mass in Ref.\cite{us31}. 

In the infrared region the self-energy is approximately constant and equal to $\mu$ up to a momentum $p\approx 3\mu$ \cite{us4}. Therefore Eq.(\ref{eq5}) and Eq.(\ref{eq6}) reflect the extreme limits, that can be obtained in a strongly interacting non-Abelian gauge theory, of how a self-energy can decrease with the momentum. Any possible self-energy solution of an asymptotically free $SU(N)$ gauge theory can be described by Eq.(\ref{eq4}) with an appropriate $\alpha$ value.
 
Eq.(\ref{eq5}) is the soft behavior dictated by OPE, while Eq.(\ref{eq6}) is the
one generated in the case where the chiral symmetry breaking is totally dominated by four-fermion interactions \cite{takeuchi,kondo}, in a limit that can also be termed as extreme walking. In the QCD case many authors claim that the asymptotic self-energy is given by Eq.(\ref{eq5}), although with $6$ flavors we are quite near the conformal window \cite{pal}. Nowadays it is known that we may have solutions with
a momentum behavior varying between Eq.(\ref{eq5}) and Eq.(\ref{eq6}) depending on the theory dynamics \cite{takeuchi,kondo}. The existence of ``effective" four-fermion interactions may change the asymptotic behavior into a hard one \cite{takeuchi,kondo}, and lattice simulations are beginning to study the self-energy behavior of $SU(N)$ theories in
the limit $\gamma \gg 1/2$. 
  
We can now discuss how the scalar masses can be computed with the help of Eq.(\ref{eq4}), and how different scalar mass values will be obtained when we vary the parameter $\alpha$ of 
that equation in the range $0$ to $1$.
The scalar mass ($m_S$) at leading order comes out from Eq.(\ref{eq2}), i.e.
\be
m_S = 2 \mu \, .
\label{eq8}
\ee
As we said $\mu$ is not an observable, and it should be written in terms of measurable quantities and by group theoretical factors of the strong interaction responsible for forming the composite scalar boson. In order to do so we will assume that the scalar composite responsible for the SM symmetry breaking give masses ($M_W$) to the electroweak bosons in the same way proposed in the traditional Technicolor (TC) models, where the dynamical mass $\mu$ is 
related to the technipion decay constant ($F_\Pi$) and to the SM vacuum expectation value ($v$) by \cite{es}
\be
M_W^2 = \frac{g_w^2 n_d F_\Pi^2}{4}  = \frac{g_w^2 v^2}{4} \, ,
\label{eq9}
\ee
where $n_d$ is the number of technifermion doublets, $v \sim 246 GeV$, and $F_\Pi$ is given by the Pagels and Stokar expression \cite{ps}
\br 
F^2_{\Pi} = \frac{N_{{}_{TC}}}{4\pi^2}&&\int \frac{dp^2p^2}{(p^2 + \Sigma^2(p^2))^2}\!\!\left[\Sigma^2(p^2) - \frac{p^2}{2}\frac{d\Sigma(p^2)}{dp^2}\Sigma(p^2)\right] \, .
\label{eq10}
\er 

Using Eq.(\ref{eq9}) and Eq.(\ref{eq10}) it is now possible to write the values of the scalar boson mass generated in a strongly interacting gauge theory in terms of the SM vacuum expectation value and the quantities that characterize a $SU(N)$ non-Abelian theory with $n_f$ fermions forming the scalar boson. The masses, calculated with the two extreme self-energy solutions giving by  Eq.(\ref{eq5}) and Eq.(\ref{eq6}) (associated to $\alpha = 0$ or $1$), are: 
\br
&& m_S^{(0)} \approx  2\left[ v \left(\frac{8 \pi^2  bg^2(2\gamma -1)}{N n_f} \right)^{1/2}\right]
\label{eq11} \\
&& m_S^{(1)} \approx 2\left[\sqrt{\frac{4}{3}} v \left( \frac{8\pi^2}{N n_{f}}\right)^{1/2}\right].
\label{eq12}
\er 
These equations involve only known quantities if we know the strongly interacting theory.
\textit{Note that the scalar boson mass, or the composite Higgs mass, scales differently with the $SU(N)$ parameters depending on the dynamics of the theory}. The factor $bg^2(2\gamma -1)$ may certainly modify mass predictions when comparing Eq.(\ref{eq11}) and Eq.(\ref{eq12}), what cannot be obtained varying naively  only $N$ and $n_f$. However, as we shall discuss in the next paragraph, the difference between the extreme values is even more complicated. Note that positivity of Eq.(\ref{eq11}) requires the constraint of Eq.(\ref{eq7}) to be obeyed.

The result of Eq.(\ref{eq8}) comes out from the comparison of the homogeneous BSE with the associated SDE \cite{es}, but \textit{the full bound state properties are subjected to the non-homogeneous BSE}, which includes its normalization condition, as clearly discussed by Llewellyn Smith~\cite{les}. 
The BSE normalization condition is given by~\cite{lane1}
\br
&2\imath&\!\!\!\left(\frac{F_\Pi}{m_{dyn}}\right)^2\!\!\! q_\mu = \imath^2\! \int d^4p \, 
Tr\left\{ {\cal P}(p)\left[\frac{\partial}{\partial q^\mu}F(p,q)\right] {\cal P}(p)\right\} \nonumber \\
&+& \int \, d^4pd^4k\, Tr\left\{{\cal P}(k)\left[\frac{\partial}{\partial q^\mu} K(p,k,q)\right]{\cal P}(p)\right\}
\label{eq13}
\er   
 where 
 \[
 {\cal P}(p)  \equiv \frac{1}{(2\pi)^2}S(p)G(p)\gamma_5 S(p),
 \]
 \[
 F(p,q) =  S^{-1}(p+q) S^{-1}(p) ,
 \]
$S(p)$ is the fermion propagator, $\Sigma (p)/\mu= G(p)$ and $K(p,k,q)$ is the BSE kernel. The BSE normalization constrain the self-energy solution if this solution is a hard one.  

Eq.(\ref{eq13}) can be divided into two parts:
\be
2\imath \left({F_\Pi}/{\mu}\right)^2 q_\nu = I^0_\nu \, + \, I_\nu^K \, . 
\label{ca11}
\ee
Contracting the above equation with $q^\nu$ and computing it at $q^2=m_S^2$, after some algebra we
verify that the final equation can be put in the form 
\be
m_S^2 \, = \, 4\mu^2 \times (I^0 \, + I^K), 
\label{eq14}
\ee
where
$I^0$ and $I^K$ are the integrals of Eq.(\ref{eq13}) contracted with the momentum $q^\nu$ and $I^K$ is a
complicated expression but of ${\cal{O}}(g^2(p^2))$ when compared to $I^0$. If we neglect the higher order term ($I^K$) we verify that \textit{the normalization condition changes the mass value given by Eq.(\ref{eq8}) by a factor $I^0$}. More importantly, $I^0$ starts being different from $1$ as we go towards the limit $\alpha \rightarrow 0$! In this limit we
obtain \cite{us1,us3}
\be 
m_S^{(0)2} \approx 4 \mu^2 \left( {3}{bg^2(2\gamma - 1)}/4{(1 + \frac{bg^2(2\gamma - 1)}{2})}\right).
\label{eq15}
\ee
First, in the limit that $\alpha \rightarrow 0$ in order to have a positive mass we recover the constraint given by Eq.(\ref{eq7}), and note that we have not yet written the factor $\mu$ in terms of observables quantities as performed to obtain Eq.(\ref{eq11}). The limit of Eq.(\ref{eq15}) can be called extreme walking, or we may also say that in this limit the chiral symmetry breaking is dominated by an effective four-fermion interaction \cite{takeuchi,kondo}. Secondly, \textit{the normalization effect lowers considerably the composite scalar masses} \cite{us1,us3,us4}.

Eq.(\ref{eq15}) has been analyzed in the case of different groups and number of fermions \cite{us1,us3}, and it was verified that it is quite difficult to obtain a light composite scalar boson associated to the SM gauge symmetry breaking, or, for instance, a scalar boson with a expected mass of  $125$ GeV when the strongly interacting theory has a characteristic mass scale
of {\cal{O}}(1)TeV. Actually we cannot say that it is impossible to obtain such light composite scalar mass, however it is quite probable that in order to obtain a light composite scalar generating the SM gauge symmetry breaking we must have a large number of fermions or fermions in a higher dimensional representation. Refs.\cite{us1,us4} contain tables and figures of composite scalar masses for different groups and number of fermions. This means that any composite scalar boson candidate to be the SM Higgs boson, when considered in isolation, will be associated to a theory with a large chiral symmetry breaking, generating a large number of Goldstone bosons. This comment will be particularly important when discussing the possibility of a light composite Higgs boson in the presence of other interactions. Of course, the result that we discussed here does not need to be linked to the SM gauge symmetry breaking, and Eq.(\ref{eq15}) tell us that a composite scalar boson, in a non-Abelian gauge theory, much lighter than its characteristic scale can be generated only for very hard asymptotic 
self-energies. 

The condition $\gamma > 1/2$ seems to be crucial to obtain a small composite scalar mass as shown in Eq.(\ref{eq15}), however this condition was not explored at depth in Ref.\cite{us1,us3}, and it is interesting
to confront this condition with lattice simulations that study  the conformal region of $SU(N)$ theories. In the case where fermions are in the fundamental representation of the
$SU(N)$ group Eq.(\ref{eq7}) implies in the following inequality
\be
\frac{9(N^2-1)}{N(11N-2n_f)}>1 .
\label{qer0}
\ee  
In the $SU(3)$ and $SU(2)$ cases this constraint indicates that we must have $n_f$ larger than $5$, and according Eq.(\ref{qer0}) the minimum $n_f$ value increases slowly as we
increase $N$. The most recent lattice data about the conformal window in $SU(3)$, which may lead to small scalar masses, indicates that its lower edge may be between 
$n_f=6$ and $n_f=8$ \cite{pal}, whereas in the $SU(2)$ this value may probably be around $6$ \cite{kara,haya}, although if this result still holds with better approximations
is still unclear \cite{appel}. 

It should be remembered that the constraint of Eq.(\ref{eq7}) is obtained at leading order, coming out from
the BSE normalization condition. Following Ref.\cite{appel1} it is reasonable to guess that the next order calculation
would modify $\gamma$ according to
\be
\gamma \approx \frac{3c16}{\pi^2 [b+(\alpha_s (\mu)/4\pi) b_1]},
\label{ga02}
\ee
where $b_1$ is the two-loop coefficient of the $\beta$ function, and to obtain numerical estimates we can assume that the strong
coupling $\alpha_s (\mu)$ has moderate values at the scale of dynamical fermion mass generation \cite{freezing1,freezing2,freezing3,freezing4}.
With this approximation the constraint $\gamma > 1/2$ can lead us even more inside the conformal window.
In any case the condition for the BSE normalization, at least for small $N$, seems not to be much different from the lattice results and even from other theoretical discussions about the conformal window of $SU(N)$ strongly interacting gauge theories \cite{ds}.

\section{Scalar composite boson mass in isolation: Effective potential approach}

The results discussed above, within the BSE approach, can also be obtained in the context of the effective potential for composite operators proposed by Cornwall, Jackiw and Tomboulis (CJT) \cite{cjt}, and a detailed calculation of the scalar composite mass through this method can be found in Ref.\cite{us5}. We stress that the same results can be obtained with the CJT potential because this potential was built just to reproduce
the SDE at the minimum, and it is not unexpected that within reasonable approximations we obtain the SDE or the BSE results. We will detail and add some points that have not been described in Ref.\cite{us5}, which turned out to be important with the discovery of the ``light" Higgs boson at the LHC.

One possibility to have a light scalar particle, no matter fundamental or composite, is when there is a symmetry protecting the boson to obtain a large mass. One clear example of such case happens when the effective Higgs potential has a classical scale invariance as proposed by Bardeen many years ago \cite{bard}, i.e. there is no $\phi^2$, or mass term, in the effective potential. Recent examples and references to earlier work on this line can be found in Ref.\cite{fin}. The first point that we would like to argue is that
\textit{a theory generating a composite Higgs boson naturally should not contain a quadratic term in the effective potential}. Quadratic terms already do not appear in the effective potential derivation of Ref.\cite{us5}, although they have been introduced in the discussion of effective theories of composite Higgses as reviewed in Ref.\cite{bella}. We will demonstrate this fact and all the argumentation follows from the effective potential for composite operators calculation \cite{cjt}.

\begin{figure}
\setlength{\epsfxsize}{0.5\hsize} \centerline{\epsfbox{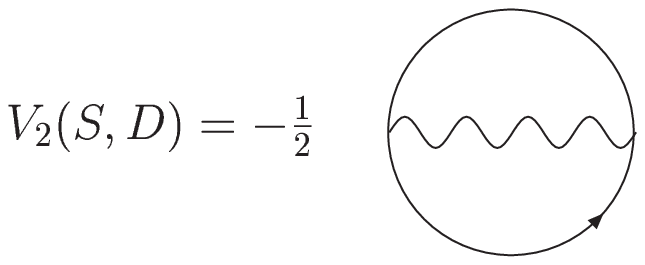}}
\caption[dummy0]{Two-particle irreducible contribution to the vacuum energy.} \label{fig1}
\end{figure}

The effective potential for composite operators can be written as
\br
V(S,D) &=& - i \, \int \, \frac{d^4p}{(2\pi)^4} \, Tr \left( \ln S_0^{-1} S - S_0^{-1} S + 1 \right) +  V_2 (S,D) \,\,\, ,
\label{eq16}
\er
where $S$ and $D$ are respectively the complete fermion and gauge boson propagators,
whereas $S_0$ and $D_0$ indicate their bare parts. Note that we are not writing the Lorentz indices and momentum integrals. In Eq.(\ref{eq16}) $V_2 (S,D)$ is the sum of all two-particle irreducible vacuum diagrams, shown in Fig.(\ref{fig1}), and can be analytically represented by
\be
i V_2 (S,D) = -\frac{1}{2} \, Tr (\Gamma S\Gamma S D) \,\,\, , 
\label{eq17}
\ee
where $\Gamma$ is the fermion proper vertex. The most important property of this potential is that its minimization with respect to the complete propagators ($S$ or $D$) lead to the 
fermionic or bosonic SDE \cite{cjt}, i.e. the minimum conditions
\be
\frac{\partial V}{\partial S}=\frac{\partial V}{\partial D}=0 \, ,
\label{vmin}
\ee
reproduce exactly the SDE for the complete $S$ and $D$ propagators.

The physically meaningful quantity that at end is going to be related to a possible effective theory is the vacuum energy density
given by
\be
\Omega_V = V(S,D) - V(S_0, D_0) \,\,\, .
\label{eq18}
\ee
The fermion propagator in terms of the self-energy is $S^{-1} = S_0^{-1} - \Sigma$, 
where $S_0 = i / \not\!p$. Expanding Eq.(\ref{eq18}) in powers of $\Sigma$ we obtain \cite{cjt,natale}
\br
&&\Omega_V = i Tr \ln (1- \Sigma S_0) + \frac{1}{2} i Tr \Sigma S_0 \Sigma S_0 +\frac{1}{2} i Tr S_0\Sigma S_0 \Sigma S_0\Gamma S_0 \Sigma S_0 \Sigma S_0 \Gamma D_0 + {\cal{O}}(\Sigma^6) \nonumber \\
\label{eq19}
\er
In the case of massless fermions $\Omega_V$ has no contribution of terms proportional to $\Sigma^1 , \Sigma^2 , \Sigma^3$, what can be verified expanding $\Omega_V$
in powers of the self-energy $\Sigma$. 

It is important to recall that taking into account the structure of the real vacuum, the propagator is a fermion bilinear
and $\Sigma$ is a gap equation depending on two momenta, i.e. $\Sigma \Rightarrow \Sigma (p,k)$. Supposing  
that the expectation value of the fermion bilinear has the following operator expansion \cite{cs}
\be
\langle \Omega | T [\chi (x+\frac{1}{2} y)
\psi (x-\frac{1}{2} y) ] |\Omega \rangle \,\,\, {}^{\,\, \sim}_{y \rightarrow 0}\,\,\,  C(y) \phi (x) ,
\label{eq20}
\ee
where $C(y)$ is a $c$-number function, and $\phi (x)$ acts like a dynamical effective scalar field. According to this
we can expect that the gap equation may be approximated by
\be
\Sigma (p,k) \sim \phi (k) \Sigma (p) \, ,
\label{eq21}
\ee
where $\Sigma (p)$ is related to the Schwinger-Dyson equation and $\phi (k)$ an effective field that is going to appear
in the effective action as a variational parameter whose minimum will be indicated in the following by $\phi$, corresponding
to the leading contribution of its expansion around $k=0$.
Details of this approach can be seen in Ref.\cite{cs}, but the point that we want to call attention next is what happens to
the term proportional to $\Sigma^2$.

The $\Sigma^2$ contribution to $\Omega_V$,for instance in the $SU(3)$ case, now indicating the momenta integrals, is proportional to
\br
\Omega_V^{(2)} &\propto & \phi^2 \int \frac{d^4p}{2\pi^4} \frac{2 \Sigma (p)}{p^2}\left\{\Sigma (p) - 3g^2 \int \frac{d^4k}{2\pi^4}\frac{\Sigma (k)}{k^2(k-p)^2} \right\} \, .
\label{eq22} 
\er
In the above expression we can recognize the fermionic SDE between brackets:
\be
\Sigma (p) = 3g^2 \int \frac{d^4k}{2\pi^4}  \frac{\Sigma (k)}{k^2(k-p)^2}  \, ,
\label{eq23}
\ee
therefore $\Omega_V^{(2)}$ is \textit{identical to zero} as long as we do have a self-energy $\Sigma (k)$ that minimizes the vacuum energy. Note that
the $\Sigma^2$ dependence of $\Omega_V$ has been obtained as an expansion of the potential in powers of $\Sigma$, and the $\Sigma^2$ disappears just
if $\Sigma (p)$ is the actual solution of the linear SDE, although the full potential is minimized by the complete non-linear SDE solution.

It may be argued that in ordinary calculations, if we work with a poor approximation for the self-energy, the equality of Eq.(\ref{eq23}) may not hold, and the final result of $\Omega_V^{(2)}$ is totally dependent on how reliable is the ansatz for $\Sigma (k)$ in the full range of momenta. \textit{For realistic theories, corresponding to a true minimum of energy, the cancellation in Eq.(\ref{eq22}) is exact, and
the effective theory must not have a $\phi^2$ term}. We should not introduce a quadratic mass term into an effective composite scalar Lagrangian only due to our inability of computing the actual self-energy. Of course, if we just work with an effective Lagrangian
to describe the composite theory, it is clear that a polynomial in terms of the effective composite scalar field starting at $\phi^4$ can always be approximated by a potential starting at $\phi^2$.

We have just argued that a theory leading to composite scalars has effective terms like $\phi^2$ suppressed. As we discuss since
the beginning we are considering all possible behaviors of the fermionic self-energies, appearing when $\alpha$ assume values from
$0$ to $1$, and when discussing the scalar mass determination in the context of the Bethe-Salpeter equation it was stressed how important is the wave function normalization, and how this condition is responsible for diminishing the scalar mass depending on the asymptotic behavior of the self-energy.
We can now discuss how this effect appears in the effective potential approach. As commented in Refs.\cite{us5}, depending
on the theory dynamics, when the self-energy decreases slowly with the momentum, i.e. in the case where $\alpha$ approaches zero in Eq.(\ref{eq4}), it is of fundamental importance the calculation of the effective Lagrangian kinetic term. 

\begin{figure}
\setlength{\epsfxsize}{0.7\hsize} \centerline{\epsfbox{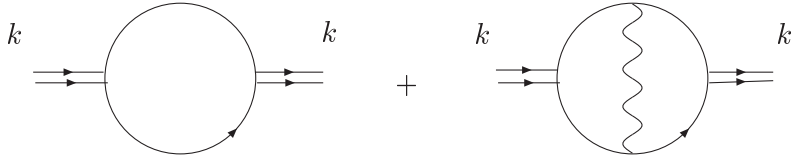}}
\caption[dummy0]{Diagrams contributing to the kinetic term in the effective Lagrangian.} \label{fig2}
\end{figure}

The kinetic term is given by the polarization diagrams ($\Pi (k^2,\phi)$), shown in Fig.(\ref{fig2}). 
In the effective action the kinetic term is the one whose denominator has the smallest power of the momentum, therefore being
the most dependent on a self-energy slowly decreasing with the momentum. It gives origin in the effective Lagrangian to a contribution
that looks like
\be
\Omega_K = \frac{1}{2}\partial_\mu \phi \partial^\mu \phi \, ,
\label{eq24}
\ee
however when the diagrams of Fig.(\ref{fig2}) are calculated it is possible to see that this term will
be multiplied by a non-trivial function ($Z$) of the number of colors of the underlying theory ($N$), the number of fermions ($n_f$), $\beta =bg^2$ and $\gamma$.
This non-trivial function, that must normalize the scalar composite field, is obtained from \cite{cs}
\be
Z \approx 2 \left. \frac{d\Pi (k^2,\phi ) }{dk^2}\right|_{k^2 =0}  \,\,\, ,
\label{eq25}
\ee	
and the $Z$ expansion around $k^2=0$ gives
\be
Z \approx \left. \frac{k^2}{8} g_{\gamma\delta} \frac{\partial}{\partial k_\gamma}\frac{\partial}{\partial k_\delta}\Pi (k^2,\phi )\right|_{k^2 \approx 0}  \,\,\, ,
\label{eq26}
\ee	
which after some algebra can be written as \cite{us5}
\be 
(Z^{(\alpha)})^{-1} \approx \frac{N n_{f}}{4\pi^2}\int dp^2\frac{(p^2)^2{\Sigma}^2(p^2)}{(p^2 + \Lambda^2)^3} \,\,\, ,
\label{eq27}
\ee 
where the index $\alpha$ is the parameter appearing in Eq.(\ref{eq4}). 

Eq.(\ref{eq27}) has been calculated in Ref.\cite{us5} in the limits $\alpha =0$ and $\alpha =1$ and the results are equal to
\be
Z^{(0)}  \approx \frac{4 \pi^2 \beta (2\gamma -1)}{N{n_f}}\left[ 1+\frac{\alpha}{\beta (\gamma -1)}+ \, ...\right] \,\,\, ,
\label{eq28l}
\ee
when $\alpha \rightarrow 0$ and
\be
Z^{(1)} \approx \frac{8\pi^2}{Nn_{f}}\left[1 - \frac{\beta\gamma}{\alpha} + ...\right] \,\,\, ,
\label{eq29}
\ee
for the case $\alpha \rightarrow 1$. 

The effective Lagrangian for the composite scalar bosons can be written as
\be
\Omega^{(\alpha )}= \Omega_K^{(\alpha )} - \Omega_V^{(\alpha )} \, ,
\label{eq30}
\ee
where, making explicit the different contributions in terms of the variational field $\phi$, we have 
\br
\Omega^{(\alpha )}=\int d^4x\left[\frac{1}{2Z^{(\alpha )}(\phi)}\partial_{\mu}\phi\partial^{\mu}\phi-\frac{\lambda_{4V}^{(\alpha)}}{4}\phi^4 - \frac{\lambda_{6V}^{(\alpha)}}{6}\phi^6 - ...\right]  \,\,\, .
\label{eq31}
\er
Eq.(\ref{eq31}) differs from an ordinary scalar field Lagrangian by the factor that multiplies the kinetic term, therefore the final effective Lagrangian comes out when we normalize the scalar field according to   
\be
\Phi\equiv [Z^{(\alpha )}]^{-\frac{1}{2}}\phi \, ,
\label{eq32}
\ee
leading to the following normalized Lagrangian
\br
\Omega^{(\alpha )}_{R} =\int d^4x\left[\frac{1}{2} \partial_{\mu}\Phi\partial^{\mu}\Phi - \frac{\lambda_{4VR}^{(\alpha)}}{4}\Phi^4 -  \frac{\lambda_{6VR}^{(\alpha)}}{6}\Phi^6 - ...\right] .
\label{eq33}
\er

The final effective theory for scalar composite fields can be written in the different $\alpha$ limits, i.e. $\alpha\rightarrow 0$ and $\alpha\rightarrow 1$, whose normalized
couplings ($\lambda_{nVR}^{(\alpha)}$) are given respectively by \cite{us5}
\br 
\lambda^{(0)}_{4VR} \equiv \lambda_{4V}^{(0)} [Z^{(0)}]^2 &\cong& \frac{Nn_{f}}{4\pi^2}[Z^{(0)}]^2\times  \left[\left(\frac{1}{\beta(4\delta - 1)} +\frac{1}{2}\right) \right. \nonumber \\
&-& \left. \frac{4\alpha}{\beta(4\delta - 1)}\left(\frac{1}{(4\delta - 2)} +2\delta\right)\right]
\er
\br 
&& \lambda^{(0)}_{6VR} \equiv \lambda_{6V}^{(0)}[Z^{(0)}]^3 = - \frac{Nn_{f}}{4\pi^2}\frac{[Z^{(0)}]^3}{\Lambda^2}  \,\,\, , 
\label{eq34}
\er
\par 
and 
\br 
\lambda^{(1)}_{4VR}\equiv\lambda_{4V}^{(1)} [Z^{(1)}]^2 \cong \frac{Nn_{f}}{4\pi^2}[Z^{(1)}]^2\left[\frac{1}{4}\left(1  + \frac{c\alpha_{{}_{S}}}{2\pi}\right)-\frac{\beta}{4\alpha}\left(\delta + \frac{c\alpha_{{}_{S}}}{8\pi}(4\delta + 1)\right)\right]\nonumber\\
\er
\br
&& \lambda^{(1)}_{6VR} \equiv \lambda_{6V}^{(1)} [Z^{(1)}]^3= - \frac{Nn_{f}}{4\pi^2}\frac{[Z^{(1)}]^3}{7\Lambda^2} \,\,\, ,
\label{eq35}
\er
where $\alpha_S$ is the coupling constant of the strong (or technicolor) interaction that forms the scalar composite, and we have to 
consider $\delta \equiv \gamma\cdot \cos ({\alpha \pi})$
expanded near the limits $\alpha =0$ and $\alpha = 1$.

\begin{figure}[h]
\centering
\includegraphics[width=0.7\columnwidth]{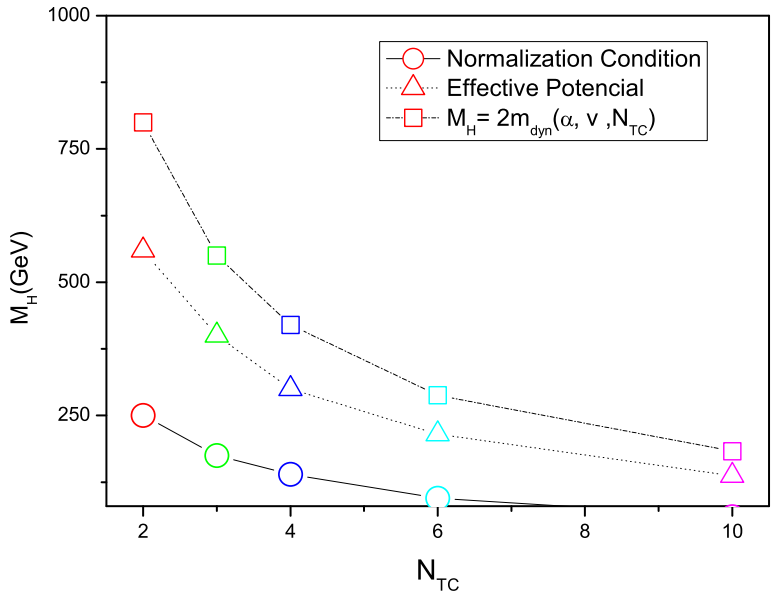}
\caption[dummy0]{ Scalar masses computed in the case of strongly interacting $SU(N)$ gauge theories with fermions in the fundamental representation. The curves were computed using respectively Eqs.(\ref{eq8}),(\ref{eq15}),(\ref{eq36}), and we used the following number of fermions : $n_f = 8,11,14,18$ for $SU(2)$ to $SU(5)$ respectively.}
\label{fig3}
\end{figure}

The scalar mass obtained from the effective potential (Eq.(\ref{eq33})) at the minimum is 
\be 
m^{2(\alpha)}_{S} \approx 2\frac{[\lambda^{(\alpha)}_{4VR}]^2}{\lambda^{(\alpha)}_{6VR}} \,\,\, .
\label{eq36}
\ee
In Fig.(\ref{fig3}) we show the differences in the scalar mass calculation using  different approaches: BSE, effective potential and
QCD like estimate (i.e. Eq.(\ref{eq8})). We consider strongly interacting $SU(N)$ theories with a large number of fermions.  The number of fermions was chosen  to be near a critical number, that for each TC group, is the  number of fermions on the border of the conformal window \cite{ds, us31}
$$
n_{f} \approx \frac{2N}{5}\frac{(50N^2 - 33)}{(5N^2 - 3)}, 
$$
\noindent where  $n_f = 8,11,14,18$ respectively for $SU(2)$ to $SU(5)$. The differences between the scalar masses computed within the BSE and the effective potential approaches are smaller than $45\%$ for small $N$ and decreases as $N$ is increased. The difference can be credited to the approximations performed when computing the effective potential for composite operators, being the crudest approximation the introduction of a very simple self-energy ansatz like the one of Eq.(\ref{eq4}) into the loop expansion, remembering that this ansatz may be reasonable
at small and large momenta, but certainly a crude approximation at intermediate momenta.

We conclude saying that, no matter we compute the scalar composite mass within the Bethe-Salpeter or in the effective potential approach, we obtain consistently small scalar masses only in the limit of a quite hard (in momentum space) self-energy. The scalar mass scales differently as we vary the dynamics of the theory, i.e. $N$, $n_f$, and $c$. Nowadays there are many reasons \cite{takeuchi,kondo,us3,us12,frezz} to believe that the self-energy may vary approximately as Eq.(\ref{eq4}) when we change the parameter $\alpha$ from $0$ to $1$. Moreover, the effective Lagrangian describing composite scalars fields should not contain a quadratic term in the scalar field if we start from a strongly interacting theory with massless fermions, contrarily to what is assumed in many discussions about composite Higgs physics. As we commented in the introductory section of this  work we consider the problem of generating one light composite scalar in a strongly interacting non-Abelian gauge theory assuming some mechanisms developed during the recent years, in particular in the Sections 2 and 3 we described in detail the main mechanisms developed for generating at least one light composite scalar in a theory with a unique characteristic scale $\Lambda$. However, the mass of a composite Higgs boson can be modified by effects inherent to the nature of the strong interaction, as, for example, due to a strong mixing between the scalar mesons as could be expected to occur in QCD,  or modified in the presence of other interactions that are not  responsible for forming the composite scalar state. In order to evaluate the implications of these possibilities and how they could change the previous results we will present in Sections 4 and 5 a brief discussion  about such possibilities. 

\section{A note about the mixing with techniglueballs and heavier scalars}

In a strongly interacting  gauge theory with $n_f$ massless fermions in the fundamental representation we have a $SU(n_f)_L\times SU(n_f)_R \times U(1)$ global chiral symmetry, which is going to be spontaneously broken by a bilinear condensate, as happens in Eq.(\ref{eq1}) in the case of the quark condensate. For a large number of fermions we shall have a large number of scalars and ($n_f^2-1$) pseudoscalars corresponding to the $SU(n_f)$ broken generators. Even in isolation there is no symmetry to impede the different scalars to mix, and this
may modify the mass predictions discussed in the previous subsections. This would be the most naive way to obtain a small scalar 
mass if we have an appropriate mixing between the different scalars. However, this may be also the most difficult problem if we consider
that this mixing is not even fully understood in the case of the QCD scalar spectra, and it is exactly based on what is known about
this spectra, that we make a few comments on the possibility of obtaining a light scalar in any ordinary strongly interacting gauge
theory.  

Supposing that we have two ordinary scalars in the theory, namely $\phi_1$ and $\phi_2$, the most simple mixing term in an effective Lagrangian would be
\be
{\cal{L}} \propto \frac{\epsilon^2}{2} \left[ \phi_1 \phi_2 + h.c. \right] \, .
\label{eq37} 
\ee
Assuming $\phi_1$ the lightest scalar formed by a fermion-anti-fermion pair, we can now suppose that $\phi_2$ may be a heavier scalar composite. This heavier composite may also be formed with a fermionic pair, but we cannot exclude the possibility that this one is the lightest scalar formed by strongly interacting gauge bosons, or what we may call a techniglueball. It is straightforward to verify that the $\phi_1$ mass, supposed to be the one computed according to the approaches discussed previously, will be modified only if we have a large mixing, i.e. a large value for $\epsilon^2$ in Eq.(\ref{eq37}). 

In the QCD case, the mixing of the QCD ``Higgs" composite, i.e. the $\sigma$, with heavier $q{\bar{q}}$ scalars and with glueballs has been
voiced for many times, see for instance Ref.\cite{mix1,mix2,mix3,mix4,mix5,mix6}. The present status of light composite scalar mesons in QCD can be also seen in the note by Amsler et al. in Ref.\cite{pdg}. A strong mixing between the scalar mesons, and in particular the $\sigma$ and the $0^{++}$ lightest
glueball has not been observed in lattice simulations \cite{lat}, although these results still need further improvements. Therefore, we
cannot foresee any particular reason for $SU(N)$ groups, different from $SU(3)$ and originating a composite scalar boson, to have a large 
scalar mixing that could change the scalar mass values ($m_S$) discussed up to now. Even if there is mixing among scalars in QCD, it is
clear that no matter is the underlying mechanism it does not lead to scalar masses lower than the QCD characteristic scale.  

\section{Scalar composite boson in the presence of other interactions}

The mass of a composite Higgs boson is modified in the presence of other interactions, i.e. any interaction that is not the one responsible for forming the composite scalar state. The prototype of such case has been discussed in Ref.\cite{foadi}, where the SM is added to a technicolor theory, and the heavy top quark mass ($m_t$) is responsible for the decreasing of the original scalar mass. This is easy to understand since the observed composite Higgs mass ($m^2_{obs}$) is the scalar mass resulting from the technicolor or
strong theory ($m^2_S$) computed above minus a one-loop fermionic correction originated from the top quark loop:
\be
m^2_{obs} = m^2_S - \kappa m_t^2 + {\cal{O}}(M_W^2,M_Z^2) \, .
\label{eq38}
\ee
In Eq.(\ref{eq38}) the positive corrections to the composite Higgs mass, proportional to the SM gauge boson masses ($M_W$ and $M_Z$), can be neglected compared to the top quark mass correction. According to Ref.\cite{foadi} the term $\kappa m_t^2$ can be large enough to reduce the scalar mass in order to have a composite Higgs boson as light as $125$ GeV.

The scenario described above works if the number of flavors of the technicolor or new strong interaction is $n_f =2$, when the breaking of the global chiral symmetry will lead only to $3$ Goldstone bosons, which are going to provide the longitudinal component to the $W$ and $Z$ bosons. In the case of $n_f >2$, when the $SU(n_f)_L\times SU(n_f)_R$ chiral symmetry is broken down to $SU(n_f)$, there are going to remain $n^2_f -4$ Goldstone bosons or technipions. Actually, these bosons are going to be pseudo-Goldstone bosons, since they will acquire mass in a realistic extended technicolor theory. Technipion masses in several models are already excluded by LHC data up to twice the top quark mass \cite{chi}. Massive technipions contribute to the scalar mass, and they contribute at the loop level to the mass given by Eq.(\ref{eq38})!
About these corrections we can make a comparison with QCD: These are the ``massive" pion loop contributions to the $\sigma$ meson mass. For QCD in isolation these corrections are zero, since the pions are massless. As we consider current quark masses the pions obtain a small mass
and the $\sigma$ meson mass has small corrections due to the small $u$ and $d$ quark masses. Of course, we are considering
that the $\sigma$ has a minor contribution from heavier quarks and mixing with glueballs. A discussion about loop corrections to the
$\sigma$ mass can be seen in Ref.\cite{nat}.

Let us now turn to technicolor with  $n_f >2$. In this case Eq.(\ref{eq38}) will be transformed into
\be
m^2_{obs} = m^2_S - \kappa m_t^2 + \xi m_\Pi^2 + {\cal{O}}(M_W^2,M_Z^2) \, ,  
\label{eq39}
\ee
where $m_\Pi$ is the technipions (or pseudo-Goldstone) masses. The determination of $\xi$, and even its signal, in Eq.(\ref{eq39}) is quite model dependent. It will involve the scalar self-couplings and the scalar technipions coupling, what may not be necessarily small numbers just based on the linear sigma model type of calculation \cite{nat}. Assuming that the scalar system is described by a linear sigma model, and $\xi$ will be of the order of $g_{\sigma\pi\pi}$, which is not a small number \cite{nat}, and assuming the model dependent limit of Ref.\cite{chi} we can estimate
\be
m^2_{obs}\geq  m_t^2 \, .
\ee
If we just imagine that the observed Higgs boson should be a composite particle, our naive estimate
indicates that the presence of the heavy top quark may not be enough to lower the composite scalar mass to the observed value.

There is another quite important fact that is missing in the effective Lagrangians when SM fermions, as the top quark, are introduced into the calculations: They generate an effective $\Phi^3$ interaction in the effective potential for composite operators. This was observed for the first time in the QCD effective Lagrangian in Ref.\cite{bar}, and in technicolor models this contribution was calculated by us \cite{us5} without a thorough analysis of the consequences. 

\begin{figure}
\setlength{\epsfxsize}{0.3\hsize} \centerline{\epsfbox{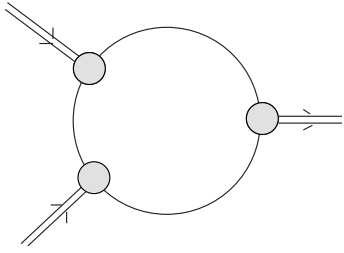}}
\caption[dummy0]{Heavy ordinary fermions ($f$ - single line) contribution to the trilinear composite ($\phi$ - double line) Higgs boson coupling. The gray blobs are proportional to the effective $ff\phi$ coupling.} 
\label{lamb3}
\end{figure}

The $\Phi^3$ contribution to the effective Lagrangian due to the
ordinary massive fermions is given respectively by the diagram of Fig.(\ref{lamb3}), where the effective $ff\phi$ coupling
is determined through Ward identities as discussed in Refs.\cite{soni,soni2,doff3}, and such coupling will be given by
\be
\imath \lambda_{\phi ff} \propto -\imath \frac{g_W \Sigma_f (k)}{2M_W} \,\,\, .
\label{eq28o}
\ee
where $g_W$ is the weak coupling and $\Sigma_f (k)$ is an ordinary fermion (or quark) self-energy. 

A full computation of ordinary fermion masses requires the introduction of an extended technicolor interaction (ETC). Without knowing the
ETC theory the best that we can do is to compute the effect of ordinary fermions to the effective potential 
as a function of their masses. The trilinear contribution to the effective potential due to the presence of ordinary fermion masses
can be computed as discussed in Refs.\cite{us5,bar}, or we can compute the trilinear coupling
as performed by Carpenter {\it et al.} \cite{soni}. Indicating the trilinear contribution to the effective
Lagrangian as $\lambda^{(0)}_{3f} \Phi^3$, we recover from \cite{us5} the largest trilinear coupling as     
\be
\lambda^{(0)}_{3f} \approx \frac{9g^3_{W}}{32\pi^2}\frac{m_{t}}{\beta(4\delta - 1)}\left(\frac{m_{t}}{M_{W}}\right)^3 \,\,\, .  
\label{eq29xx}
\ee
This is the largest contribution to the trilinear coupling, and it was obtained in the limit $\alpha \rightarrow 0$ of Eq.(\ref{eq4}). The details
of the calculation are in the appendix of Ref.\cite{us5}. 

The $\lambda^{(0)}_{3f} \Phi^3$ contribution to the effective potential is the
one that may decrease the scalar mass, and is the equivalent of the negative top quark contribution in Eq.(\ref{eq38}). However this
coupling must be compared to the quadrilinear contribution to the potential $\lambda_4 \Phi^4$ (see \cite{us5}), which has the same
effect of the technipions self-coupling present in Eq.(\ref{eq39}). It is the balance between these different couplings that will
lower or increase the scalar composite mass.

In simple words we may say that the contribution of ordinary fermions,
even considering the effect of the top quark mass in Eq.(\ref{eq29xx}) \cite{com1}, is small due to the presence of
the $g^3_W$ weak coupling and the other numerical factors,
and hardly compete with larger terms $\lambda_4 \Phi^4$ (and $\lambda_6 \Phi^6$) of ${\cal{O}}(1)$, characteristic of the composite
scale ($\Lambda_{TC}$), and which are responsible for the determination of the composite scalar boson mass. A precise calculation must involve specific models, and at this level we may
say the the top quark effect may lower the composite scalar mass, but at the cost of precise cancellation of different terms in the effective Lagrangian. The most we can learn
from this discussion is that with a heavy top quark, and \textit{in the presence of massive technifermions, the trilinear terms in the effective composite scalar Lagrangians cannot be always neglected in the study of phenomenological models}. 

\section{Light composite scalar boson in two-scale TC models}

\par One possibility for generating a light composite scalar happens in the case where we have at least two composite bosons ($\phi_1$,$\phi_2)$  and they have a strong mixing \cite{ff}. In this case we may have a see-saw mechanism where one of the scalar composites may turn out to be quite light. The main problem along this line is to find the condition for the existence
of this strong mixing in the case where we have several composite scalar bosons. In a single theory with a unique scale, as we discussed before, we do not know of any mechanism in 
this direction. Therefore we shall consider a two-scale model where it is simple to verify why a strong mixing between scalar composite appears, and
speculate how this mechanism can be extended to other cases \cite{usx}.  

Consider the formation of a light composite scalar  boson when the TC theory features two technifermion species in different representations,  $R_1$ and $R_2$,  under a single technicolor gauge group, with characteristic scales $\Lambda_1$ and $\Lambda_2$. Walking technicolor theories can  have fermions belonging to different technicolor  representations and, therefore, may have two different scales  with  characteristic  chiral symmetry  breaking scales $\Lambda_1(R_1) < \Lambda_2(R_2)$. Here we assume that  technifermions  are in  the representations $R_1$ and $R_2$ under a single technicolor gauge group as described in Ref.\cite{Lane}.     
In the model proposed by Lane and Eichten, it is assumed that the TC running coupling constant is given by 
\br 
&&\alpha_{TC}(p^2) = \alpha_2  \,\,{\rm when }\,\,p > \Lambda_2 \nonumber  \\
&&\alpha_{TC}(p^2) = \alpha_1\left[1+ \beta^1_{0}\ln\left(\frac{p^2}{\Lambda^2_1}\right)\theta(p^2 - \Lambda^2_1) \right]^{-1} \nonumber \\
&& \hspace{2.0cm} \!\!{\rm when }\, \Lambda_1 < p < \Lambda_2 \nonumber 
\er 
 \noindent where $\alpha_2 = \alpha(R_2) = \frac{\pi}{3C_2(R_2) }$, $\alpha_1 = \alpha(R_1) = \frac{\pi}{3C_2(R_1) }$ \cite{sus}  and  $\beta^1_{0} =  \frac{\alpha_1}{6\pi}(11N_{TC} - 4N_1)$ and $N_1$ are  technifermions doublets in the representation $R_1$.  Note that the $N_1$  and $N_2$ doublets of technifermions   belong  to the complex TC representation $R_1$ and $R_2$,  with  dimensionality $d_1 < d_2$. For a large enough  number of $N_1$  doublets it is possible to obtain $\Lambda_1(R_1) << \Lambda_2(R_2)$ \cite{Lane} (or the decay constant $F_1 << F_2$) because 
\be 
\frac{\Lambda_2}{\Lambda_1} \approx \exp{\left(\frac{6\pi}{(11N_{TC} - 4N_1)}\left[ \alpha^{-1}(R_2) - \alpha^{-1}(R_1) \right]\right)}, 
\ee 
\noindent in this case we can assume that the asymptotic technifermions self-energy behavior in  representation $R_1$  can be described  by Eq.(4), this hypothesis can be verified with the numerical results obtained in \cite{Lane}. 


\par  At low energies  we have an effective theory containing two different sets of composite scalars  $\phi_1$  and $\phi_2$, like the ones described in Ref.\cite{Lane}. We will assume an  ETC gauge  group  containing $N_1$ technifermions doublets in the fundamental representation  $R_1 = F$, and $N_2$ technifermions doublets, assuming  $N_2 = 1$ for $R_2$ representations (2-index antisymmetric $A_2$, 2-index symmetric $S_{2}$). The phenomenology of these type of models was already described in Ref.\cite{Lane}.

\begin{figure}[ht]
\begin{center}
\includegraphics[width=0.7\columnwidth]{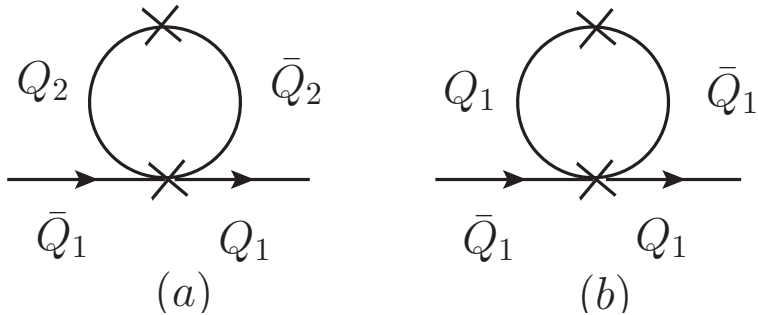}
\caption{ETC (effective four-fermion) contributions to the mixing of scalars in the representations  $R_1$ and $R_2$ }
\label{fig1x}
\end{center}
\end{figure}

The fermionic content of the  model that we discuss in this section  has two multiplets of technifermions in the representations $R_1$ and $R_2$ of the type  
\br 
&& Q^{U}_{ETC} = \left(\begin{array}{c} {U^{a}_{R_1}}_1 \\ \vdots \\ {U^{a}_{R_1}}_i \\ {U^{a}_{R_2}}_1  \end{array}\right)_{L,R}\nonumber \,\,,\,\,Q^{D}_{ETC} =\left(\begin{array}{c} {D^{a}_{R_1}}_1 \\ \vdots \\ {D^{a}_{R_1}}_i \\ {D^{a}_{R_2}}_1   \end{array}\right)_{L,R}\\ \nonumber
\er
\noindent where $(a)$ is a technicolor index and  $(i)$ is a flavor index.  In this type of theory the ETC group would be $SU(N_ {ETC})\supset SU (N_ {TC} + N_1 + N_2)$, and  in order to incorporate the mixing between  $\phi_1 $ and $\phi_2$, we must take into account the contributions of the ETC as displayed in  Fig.(\ref{fig1x}). Remembering that the self-energy can also be related to the solutions of the Bethe-Salpeter equation, we can observe that the scalar boson $\phi_1 $, formed by the fermions in the representation $R_1$ receive contributions of the condensates of the two different representations, as shown in Fig.(\ref{fig1x}).

\par We can detail a little bit more the comment of the previous paragraph and the behavior of the diagrams in Fig.(\ref{fig1x}). The $Q_1$ techniquarks will receive a dynamical mass due to the usual TC contribution and to the two diagrams in Fig.(\ref{fig1x}), that we indicate by
\be 
\Sigma_{Q_1}(p) \approx \Sigma^{TC}_{Q_1}(p)  + \zeta\Sigma_{Q_2} + \xi\Sigma_{Q_1},
\ee 
where $\zeta$ and $\xi$ are calculable constants.
In the above expression the first one is the usual TC contribution due to condensation of $Q_1$ techniquarks in the representation $R_1$. The second comes from the ETC interaction with $Q_2$ techniquarks that condensate in the representation $R_2$ and the third one is the $Q_1$ contribution  from  ETC  interactions.  Suppose now that the $Q_1$ techniquarks self-energy does not have a walking behavior, or a slow decrease with the momentum i.e. 
$\Sigma_{Q_1} (p^2)$ is given by Eq.(4), therefore the $Q_1$  ETC contribution to $\Sigma_{Q_1}(p)$,  Fig.(\ref{fig1}b) will be given by 
\cite{us3}
\be 
\xi \Sigma_{Q_1} \propto  O(\frac{\Lambda^3_{1}}{\Lambda^2_{ETC}}) << 1 \, ,
\ee 
which is totally negligible.

\par We can now consider the effect of $Q_2$ technifermions in the representation $R_2$. This contribution is represented by the diagram  
of Fig.(\ref{fig1x}a), where we may have an extreme walking behavior for the $Q_2$ technifermions. In this case the correction due to ETC will be dominated by a self-energy of the type given by Eq.(\ref{eq6}) resulting in \cite{us3}
\be 
 \zeta \Sigma_{Q_2} \approx \Lambda_2\left(\frac{C_{ETC}}{C_{2R_2}}\left(\frac{\alpha_{ETC}(\Lambda^2_{ETC})}{\alpha_{TC}(\Lambda^2_{ETC})} \right)^{\gamma_{2}}\right).
\label{eqdel2}
\ee
\noindent Therefore the ETC correction ($\zeta \Sigma_{Q_2}$)  plays a role similar of a bare mass term for the $\Sigma_{Q_1}(p)$ self-energy, i.e. a very hard self-energy! A similar reasoning may also be applied to  the $\Sigma_{Q_2}(p) \approx \Sigma^{TC}_{Q_2}(p)  
+ \kappa \Sigma_{Q_1} $. \textit{Although only one of the technifermions representations of one given TC group has a walking behavior and
this group belongs to an ETC theory, at the end both technifermions representations will have asymptotically hard self-energies}.
In the following we will consider that the technifermions associated to the representation $R_1$ are in the fundamental representation  with a 
self-energy behaving as the one of Eq.(\ref{eq5}), and  $\Sigma_{Q_2}(p) \approx \Sigma^{TC}_{Q_2}(p)$ behaving as Eq.(\ref{eq6}) . 

The argumentation about the origin of a strong mixing between the two composite scalars will be discussed in terms of the different contributions
to the CJT potential, and as will be explained in the sequence, this mixing will be enhanced only if one of the self-energies decreases slowly with the momentum.

\par The different terms that are going to appear in the effective action of the composite scalar system are momentum integrals of different powers of the 
self-energies $\Sigma (p)$ \cite{cjt}, which are going to be represented as $[\phi_i \Sigma_i (p)]^n$, where $\phi_i$ acts like a dynamical
effective scalar field (expanded around its zero momentum value) \cite{us5,cs}, and it is interesting to verify how it is going to be 
the behavior of the $\Sigma_i^4(p)$ term (as a function of the momentum), which is the leading term of the effective potential \cite{us5,cs}.

The fourth power of the self-energy associated to the fields $\phi_1$ and $\phi_2$, where the index $1$ will be related to technifermions with (in principle) a soft self-energy ($\alpha =1$), and the index $2$ will be related to technifermions in a representation $R_2 = S_2$ or 
$R_2 =  A_2$, with a hard self-energy ($\alpha =0$), will be written as
\br 
\Sigma^4_{1}(p^2) = && \hspace*{-0.3cm}\left( \Lambda_1 f(p)  +   a_{ETC}\Lambda_2 \right)^4 \approx \Lambda^4_1 f^4(p) + \nonumber \\
                    &&\hspace*{-0.3cm} 4a_{ETC}\Lambda^3_1\Lambda_2  f^3(p) + 6a^2_{ETC}\Lambda^2_1\Lambda^2_2  f^2(p)+...\label{4mix0} \\
\Sigma^4_{2}(p^2) = && \hspace*{-0.3cm} \Lambda^4_2	\left[1 + \beta_{0}(R_2)\ln\left(\frac{p^2}{\Lambda_2}\right) \right]^{-4\gamma_2} \label{4mix1}							
\er 
\noindent  where we defined $f(p) = \Lambda^2_1/p^2$ and $a_{ETC}$ is the ratio of Casimir operators and couplings of Eq.(\ref{eqdel2}).

After some lengthy calculation, that follows the same steps delineated in Ref.\cite{us5}, we obtain the following effective Lagrangian using the self-energies described previously  
\br 
\Omega(\Phi_1, \Phi_2) = \hspace*{-0.3cm}&&\int d^4x\left[\frac{1}{2}\partial_{\mu}\Phi_1\partial^{\mu}\Phi_1 + \frac{1}{2}\partial_{\mu}\Phi_2\partial^{\mu}\Phi_2  \right. \nonumber \\
&& \left. -\frac{\lambda^{R_1}_{4n}}{4}Tr\Phi_{1}^4 - \frac{\lambda^{R_2}_{4n}}{4}Tr\Phi_{2}^4 - \frac{\lambda^{R_1,R_2}_{4n}}{4}Tr\Phi_{1}^2\Phi_{2}^2 \right.\nonumber \\ 
&& \left.  - \frac{\lambda^{R_1}_{6n}}{4}Tr\Phi_{1}^6  - \frac{\lambda^{R_2}_{6n}}{4}Tr\Phi_{2}^6 \right].
\label{omfin}
\er 

\noindent The coefficients,  $\lambda^{R_i}_{4n}, \lambda^{R_i}_{6n}$ to $i=1,2$,  shown in equation  above are described in Ref.\cite{usx}.  

 The most important characteristic of this effective Lagrangian is the mixing term
\be
\lambda^{R_1,R_2}_{4} = \frac{3N_{TC}N_1}{2\pi^2}\left(\frac{C_{ETC}}{C_{2R_2}}\left(\frac{\alpha_{ETC}(\Lambda^2_{ETC})}{\alpha_{TC}(\Lambda^2_{ETC})} \right)^{\gamma_{2}}\right)^2 .
\label{mix0}
\ee
This mixing is the one that defines the splitting between the effective fields $\phi_1$ and $\phi_2$, as discussed by Foadi and Frandsen 
\cite{ff}, whereas within the approach taken in this work their parameter $\delta$ \cite{ff}, characterizing the mixing  in the mass  matrix,  is
\be 
\delta = \frac{\lambda^{R_1,R_2}_{4n}}{\sqrt{\lambda^{R_1}_{4n}\lambda^{R_2}_{4n}}}.
\label{xx} 
\ee 
We emphasize that this mixing appears naturally in a two-scale TC model, where it is enough that one of the scales, and the fermionic
representation associated to it, has an extreme walking behavior and the TC group is embedded into an ETC theory. 

\par Considering $F_{2}\sim 250GeV$,  we note that the scale $\Lambda_2 $ is defined by
\be
N_2 F^2_{2}  = \frac{\Lambda^2_2}{Z^{(0)}}
\ee 
\noindent  which leads to
\be 
\Lambda_2 = \frac{2\pi F_{2}\sqrt{\beta(2\gamma - 1)}}{\sqrt{N_{TC}}} \sim  \frac{O(TeV)}{\sqrt{N_{TC}}}. 
\ee 
\noindent Finally, assuming  
\be 
M^{2}_{\Phi_i} = \frac{\partial^2\Omega(\Phi_i)}{\partial \Phi^2_i}\left|_{\Phi = \Phi_{min}}\right.
\ee 
\noindent  we obtain
\br 
&& M^{2}_{{}_{\Phi_i}} \approx 2\lambda^{R_i}_{4n}\left(\frac{\lambda^{R_i}_{4n}}{\lambda^{R_i}_{6n}}\right). 
\er

\par We can write the following mass matrix in the base formed by the composite scalars ($\Phi_1$)  and ($\Phi_2$)  
\br
M^2_{\Phi_1 ,\Phi_2} = \left(\begin{array}{cc} M^2_1 & M^2_{12}\delta \\ \delta M^2_{12} & M^2_{2} \end{array}\right).
\label{Meig}
\er 
\noindent The eigenvalues of this matrix provide the mass spectrum for the light scalar $(H_1)$ and heavy $(H_2)$, including the 
mixing effect parametrized by $\delta$, where 
\br 
&& M^2_i = 2\lambda^{R_i}_{4n}\left(\frac{\lambda^{R_i}_{4n}}{\lambda^{R_i}_{6n}}\right) \nonumber \\
&& M^2_{12} = M_1M_2 .
\er 
\par
\begin{figure}[h]
\begin{center}
\includegraphics[width=0.5\columnwidth]{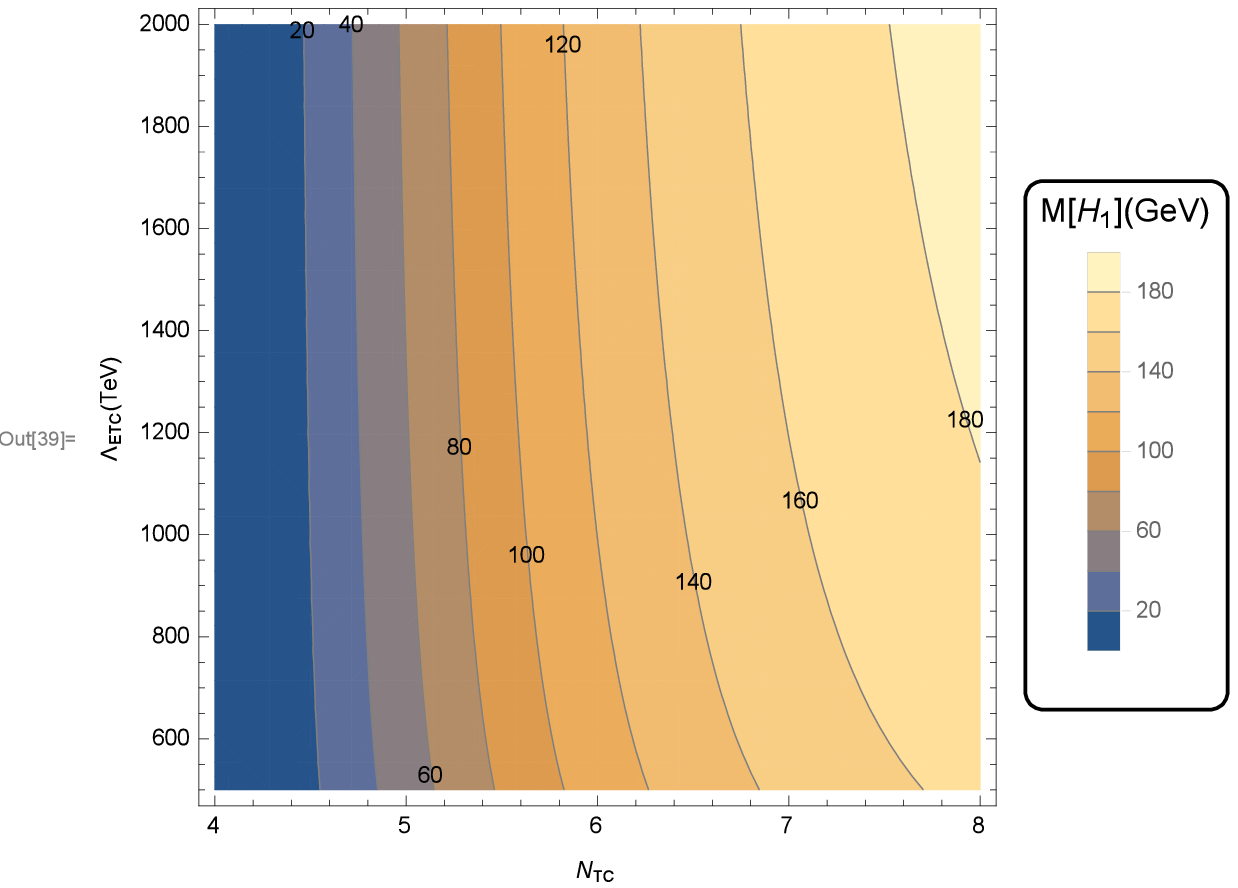}\,\includegraphics[width=0.5\columnwidth]{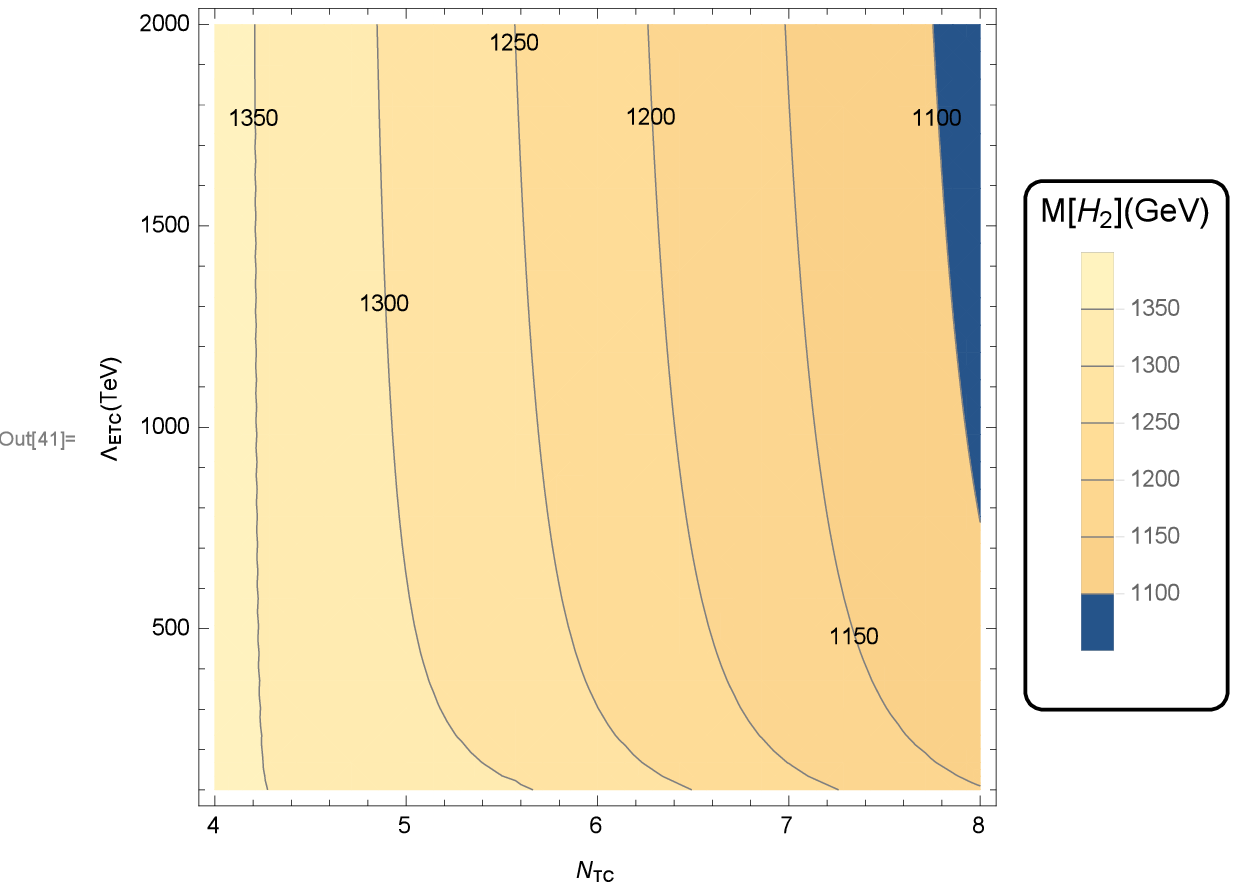} 
\caption{The light composite scalar $H_1$ and heavier composite scalar $H_2$ regions of masses as a function of the parameters $N_{TC}$ and $\Lambda_{ETC}$. }
\label{fig3b}
\end{center} 
\end{figure}

  From the above equations we can determine the mass spectrum for the scalar bosons, $H_1(R_1)$ and $H_2(R_2)$, which are the diagonalized masses of the scalars $\Phi_1$ and $\Phi_2$. As an example we present  the mass spectrum obtained for the  light  and  heavier  composite scalars $H_1(R_1)$ and $H_2(R_2)$ in the cases where $R_1=F$,  $R_2 = A_2$( Fig. (\ref{fig3b})).   

\par We computed an effective action for a composite Higgs boson system formed by two technifermion species in different representations,  $R_1$ and $R_2$,  under a single technicolor gauge group with characteristic scales $\Lambda_1$ and $\Lambda_2$ as the original proposal presented  in Ref.\cite{Lane}. The calculation is based on an effective action for composite operators\cite{us5}, 
the novelty of the calculus presented in this section is that  we included technifermions  in different representations,  $R_1$ and $R_2$,  under a single technicolor gauge group. The mixing between the composite scalar bosons  $\Phi_1$ and $\Phi_2$ can generate a light Higgs boson of mass of  approximately ${\cal{O}}(100)$GeV. To obtain a large mixing it is enough that one of the technifermions representations has a walking behavior and the TC group is embedded in an ETC theory. At the end the technifermions of both representations will have asymptotically hard self-energies.
 
For a set of parameters similar to the ones used in  Ref.\cite{Lane} in the case $R_2=A_2$, we obtain the same TC group necessary to generate the walking behavior, $SU(6)_{TC}$, leading  also to small scalar masses. 
 Furthermore, the large anomalous dimensions $\gamma_m$ enhance light-scale technipion masses, $M_{\pi1} > M_{\rho1} - M_{W} $, where technirho mass  $M_{\rho1} \sim 250GeV$.  The difference between the results obtained for the representations  $R_2=A_2$  and $R_2=S_2$(see Ref.\cite{usx}) is that in the $A_2$ case we obtain a light  scalar mass only with a large ETC scale .  For the heavy scalar bosons obtained with $R_2=S_2$ or $R_2= A_2$  we expect the mass to be in the range $[1200-1300]$ GeV. Concluding that a light composite scalar boson mass can be generated in a type of see-saw mechanism in a two-scale TC model. 
 
Note that the main reason for generating a strong mixing between the effective composite scalar fields appears in Eqs. (\ref{4mix0}) and (\ref{4mix1}), leading to the mixing term
in the effective potential (see Eq.(\ref{omfin})). If the self-energies of the different techniquarks were soft the composite scalars barely mix, however if one of
the self-energies is of the hard type it is enough to generate a strong mixing and a light scalar. Of course, if one the self-energies decreases slowly with the momentum the condition
for a light scalar existence is already in there. Looking at this it may be considered that we introduced a very complicated model just to obtain a light scalar, but it should be
remembered that we do have in this case an extra possibility to manipulate the scalar masses values, i.e. we can obtain an even larger mass splitting between different composite scalar boson masses than the one obtained in the case when we just have only one hard self-energy. 

\section{The case of many composite scalars}

It is possible that the mixing mechanism that we discussed in the previous section may be extended to models with
more than one TC group, although it is also possible to envisage that in this case we shall need a more complex ETC interaction in
order to mix the different groups. This possibility may occur under two conditions: The first one is that one of the scalar composites is formed by fermions of a given group and representation such that their self-energy is of the extreme walking type. The second one imply in a strong mixing between the different scalar composites, this will happens if the different scalar bosons
interact through an ETC generating effective four-fermion interactions like the ones displayed in Fig.(\ref{fig1x}). 

As one example, suppose that we have several TC groups, $SU_{TC1}(N_1)$, $SU_{TC2}(N_2)$,... We must have at least one of the $SU_{TCi}(N_i)$ with a walking behavior, and this one be embedded into an ETC theory together with some of the others TC groups, which may be characterized by a high energy scale and standard soft self-energy, leading to very massive composites.
The fact that they mix through ETC interactions will generate a see-saw mechanism, in such a way that at least one of the scalar composites will become quite light, when compared to the mass scales of all other TC theories. 

The most economical case will happens when the light composite states will belong to an $SU_{TC}(2)$ group, with $N_d=1$  technifermion doublet, providing exactly the degrees of freedom necessary to break the electroweak group and generating only one light scalar composite, as seems to be the experimental scenario observed up to now. The $SU_{TC}(2)$ group that we are referring about does not necessarily should have an almost conformal dynamical behavior, but it must mix with another TC group endowed with an extreme walking behavior, and this mixing has to be tuned in order to provide exactly the degrees of freedom necessary to implement the weak boson masses and generate a light scalar boson. All the other composite states may be quite heavy and not accessible to the LHC energies.
 
\section{Remark on vector composites}

There is a large amount of phenomenology describing the possibility of observing vector composites, like the technirho or techniomega. However it is interesting to recall the remark made by some of us in Ref.\cite{us4} that the vector composites are not going to be light, are not going to be modified by the dynamics that leads to light scalars, and may be difficult to observe at the LHC. The vector composites in a non-Abelian gauge theory
are quite massive basically due to the spin-spin part of the
hyperfine interactions. For $S$ waves the hyperfine splitting has been determined as \cite{ei1,ei2}
\be
M(^3S_1)-M(^1S_0)\approx \frac{8}{9} {\bar{g}}^2(0) \frac{|\psi (0)|^2}{\mu^2} \, ,
\label{eqvc}
\ee
where $|\psi (0)|^2$ is the meson wave function at the origin. We also assume that the fermion masses forming the vector boson are equal
to the dynamical mass $\mu$. Eq.(\ref{eqvc}) has been derived in the heavy quarkonium context \cite{ei1,ei2}, although it seems to
work reasonably well even in the presence of light fermions \cite{sc}.
 
Assuming as the worst possibility that ${\bar{g}}^2(0)/4\pi \approx 0.5$ what is consistent with a frozen infrared coupling constant \cite{freezing1,freezing2,freezing3,freezing4}, the fact that no other composite  scalar boson has been found below $125$ GeV, and that $|\psi (0)|^2 \approx \mu^3$,
what is consistent with Eq.(\ref{eq3}), we obtain the following inequality from Eq.(\ref{eqvc})
\be
M(^3S_1) > (2\pi \mu + 125) \mbox{GeV} .
\label{eqvec}
\ee
With the dynamical fermion mass values (or characteristic TC scales $\Lambda_i$) that we have discussed in this work, 
the vector composite masses are going to be heavy and with a quite model dependent phenomenology. The main point of this section was to recall that possibly the experimental signal
of vector composites will hardly be affected by the specific dynamics of the self-energies that we discussed up to now, and probably the ratio between scalar and vector masses
will not be helpful to reveal the underlying strong interaction dynamics. 

\section{Results and conclusions}

We discussed the possibilities for generating a light composite scalar boson in non-Abelian $SU(N)$ strongly interacting gauge theories.
In Sections 2 and 3 this problem was studied with two different approaches: The solutions of Bethe-Salpeter equations and the effective potential for composite operators. Both approaches
lead to the same result, indicating that light composite scalar bosons appear as a direct consequence  of extreme walking (or quasi-conformal) technicolor theories, where the  asymptotic self-energy  behavior  is described by Eq.(\ref{eq6})(or $\alpha =0$). 

In the BSE approach the extreme walking  behavior of fermionic self-energies and the wave function of scalar bound states are dominated by higher order interactions and are characterized by a much harder decrease with the  momentum, and  in this case the normalization condition of the BSE do constrain the scalar masses, resulting in  a light composite scalar boson. 

In the effective potential approach the normalization  constant $Z(\alpha)$ is important to set the right scale in our  effective Lagrangian, the scalar mass obtained in this approach is proportional to this constant, see Eq.(\ref{eq36}), in the extreme walking behavior ($\alpha = 0$) we obtain $Z(0) \sim O(\frac{Z(1)}{10})$ ,  where $Z(1)$ is the normalization factor calculated with $\alpha = 1$, which is associated to the known asymptotic self-energy behavior predicted by the operator product expansion (OPE). The consequence of the extreme walking behavior is the reduction of the scalar composite boson mass by almost one order of magnitude. 

In Section 3 we commented that quadratic terms are not natural in effective Lagrangians
generated through the calculation of the effective potential for composite operators, where they may appear only when poor approximations for the self-energies are used to compute the effective action. We also show that the scalar boson mass scales differently with the theory parameters (such as $N$ and $n_f$) according to the theory dynamics, i.e. according
to the different self-energies behavior.

In the Section 5 we discussed how the mass of a light composite scalar boson can be modified in the presence of other interactions, i.e. any interaction that is not the one responsible to form the composite scalar state, as an example we consider the prototype  discussed in Ref.\cite{foadi}, where the SM is added to a technicolor theory and the heavy top quark mass ($m_t$) is responsible to  decrease the original scalar mass. This case is realistic when the strongly interacting theory is a minimal one. When the strong or TC theory has a larger chiral symmetry,
there will remain many pseudo-Goldstone bosons, and they will contribute to the scalar mass with a signal contrary to the one of the top quark. In the effective potential for composite operators these new interactions will be responsible for new cubic and quartic interactions. 

In the Section 6 we considered the  possibility of generating a light composite scalar  where we have at least two composite bosons and they have a strong mixing, to obtain a large mixing it is enough that one of the technifermions representations has a extreme walking behavior and the TC group is embedded in an ETC theory. In this case we have a see-saw mechanism where one of the scalar composites turn out to be quite light, and we obtain a light scalar boson mass of approximately a few hundred GeV. In Section 7 we presented a brief discussion about how the mixing mechanism, presented in the Section 5, can be  extended to models with more than one TC group, and in Section 8 we remark 
that vector composites masses are not going to be affected by the mechanisms that help to lower the scalar mass, since most of their
masses are originated by the spin-spin interaction, and if there is any new strongly interaction at the TeV scale the vector composites
may scape detection in the near future. 

Summarizing, in this work we identified that, regardless of the approach used for generating a light composite scalar boson, the behavior exhibited by the extreme walking (or quasi-conformal) technicolor theories is the main feature needed to produce a light composite scalar boson, i.e. a mass roughly one order of magnitude below the characteristic mass scale of the strongly interacting theory responsible for its formation. It is also interesting to note, according to what was presented in Section 5 (and 6), that a strong mixing between composite scalars may appear in the presence of several strongly interacting theories, leading to a see-saw mechanism and a light
composite scalar bosons, just if one of the theories has a walking behavior, which is going to be transmitted to the other theories if
they are embedded into a large ETC theory. In all the cases it seems that a slowly decreasing self-energy with the momentum is a necessary condition for the new fermions of a strongly
interacting theory to form a light composite scalar.

\section*{Acknowledgments}

This research was partially
supported by the Conselho Nacional de Desenvolvimento Cient\'{\i}fico e Tecnol\'ogico (CNPq)  by grants 442009/2014-3 (AD) and 302884/2014-9 (AAN), by the Coordena\c c\~ao de Aperfei\c coamento de Pessoal de N\'{\i}vel Superior (CAPES) (AAN), and by Funda\c c\~ao de Amparo
\'a Pesquisa do Estado de S\~ao Paulo (grant 2013/22079-8).


\begin{thebibliography}{99}
\bibitem{com1} Note that the enhancement due to the top quark mass that we refer here is how its large mass acts in the sense of decreasing the scalar composite mass.
\end{thebibliography}
\end{document}